\documentclass[preprint, superscriptaddress, amsmath,amssymb, aps, prb]{revtex4-1}
\usepackage{graphicx}% Include figure files
\usepackage{dcolumn}% Align table columns on decimal point
\usepackage{bm}% bold math
\usepackage{amsmath}
\usepackage{epsfig}
\usepackage{rotating}
\usepackage{enumerate}
\usepackage{xcolor}

\def\etal{{\em et al.\/}}

\def\eva3{$eV$$\cdot$\AA$^3$}               % eV.A^3
\def\*#1*/{}                        % Allows commenting out whole paragraphs

\newcommand{\degree}{\ensuremath{^\circ} }

\begin{document}

\title{A phononic switch based on ferroelectric domain walls}

\author{Juan Antonio Seijas-Bellido}
\affiliation{Institut de Ci\`{e}ncia de Materials de Barcelona (CSIC), Campus de Bellaterra,
08193 Bellaterra, Barcelona, Spain}

\author{Carlos Escorihuela-Sayalero}
\affiliation{Materials Research and Technology Department, Luxembourg Institute of Science and Technology (LIST), 
5 avenue des Hauts-Fourneaux, L-4362 Esch/Alzette, Luxembourg}

\author{Miquel Royo}
\affiliation{Institut de Ci\`{e}ncia de Materials de Barcelona (CSIC), Campus de Bellaterra,
08193 Bellaterra, Barcelona, Spain}

\author{Mathias P. Ljungberg}
\affiliation{Donostia International Physics Center, Paseo Manuel de Lardizabal, 4, 
E-20018 Donostia-San Sebasti\'an, Spain}

\author{Jacek C. Wojde{\l}}
\affiliation{Institut de Ci\`{e}ncia de Materials de Barcelona (CSIC), Campus de Bellaterra,
08193 Bellaterra, Barcelona, Spain}

\author{Jorge \'I\~niguez}
\affiliation{Materials Research and Technology Department, Luxembourg Institute of Science and Technology (LIST),
5 avenue des Hauts-Fourneaux, L-4362 Esch/Alzette, Luxembourg}

\author{Riccardo Rurali}
\affiliation{Institut de Ci\`{e}ncia de Materials de Barcelona (CSIC), Campus de Bellaterra,
08193 Bellaterra, Barcelona, Spain}

\email{rrurali@icmab.es}

\date{\today}

\begin{abstract}
The ease with which domain walls (DWs) in ferroelectric materials
can be {\em written} and {\em erased} provides
a versatile way to dynamically modulate heat fluxes.
In this work we evaluate the thermal boundary resistance (TBR) of 
180\degree DWs in prototype ferroelectric perovskite PbTiO$_3$ within the numerical formalisms of 
nonequilibrium molecular dynamics and nonequilibrium
Green's functions. An excellent agreement is obtained for the TBR of an isolated DW derived from 
both approaches, which reveals the harmonic character of the phonon-DW scattering mechanism.
The thermal resistance of the ferroelectric material is shown to increase up to around 20\%,
in the system sizes here considered, due 
to the presence of a single DW, and larger resistances can be attained by incorporation of more 
DWs along the path of thermal flux. These results, obtained at device operation temperatures,
prove the viability of an electrically actuated phononic switch based on ferroelectric DWs.
\end{abstract}

\maketitle


% Introduction

The modulation of the thermal flux is necessary to encode basic logic 
functions in devices that operate with heat currents, rather than with charge
carriers or electromagnetic waves. This task, however, has thus far been elusive.
The reason, simply put, is that phonons, the quantized vibration
of the lattice that carry heat in an insulator, have no mass and no 
bare charge and therefore their motion cannot be easily controlled with
an external field~\cite{LiRMP12}.

A few configurations, based on heterojunctions~\cite{DamesJHT09,
RuraliPRB14} or structural asymmetry~\cite{YangAPL08,CartoixaNL15},
have been proposed to implement thermal diodes~\cite{TerraneoPRL02,
LiPRL04,RobertsIJTS11}, where a preferential direction for heat 
transport exists, i.e. the thermal conductivity depends on the sign 
of the thermal gradient. A step further, however, would be the design of 
structures where the thermal conductivity can be {\em dynamically} 
tuned. In particular, a system able to reversibly commute between 
a {\em low} and a {\em high} conductivity state would pave the way 
to digital signal processing with phonons. 

Perovskite ferroelectric oxides provide in principle an effective 
way to pursue this goal. These materials have a spontaneous 
electric dipole moment determined by the off-center displacement of
the cations with respect to the surrounding oxygen cages. Such a 
polarization can be reoriented or fully reversed with an external 
electric field, and configurations consisting of juxtaposed domains 
with different polarization can be designed. Therefore, domain walls 
(DWs) that separate uniformly-polarized domains can be dynamically and 
reversibly written and erased. In this work we explore to
what extent a specific type of PbTiO$_3$ DWs, the so-called
180\degree DWs occurring between adjacent domains of antiparallel 
orientation of the ferroelectric polarization, act as scattering 
planes for the incoming phonons, thus yielding a thermal boundary 
resistance (TBR) that can be switched on and off with an external 
electric field.

A proof-of-concept of this idea has been experimentally observed in
different ferroelectrics and DW configurations. Mante \etal~\cite{Mante1971} first and
Weilert \etal~\cite{Weilert1993,Weilert1993a} later, demonstrated that 
the thermal conductivity of bulk BaTiO$_3$ and KH$_2$PO$_4$
can be dynamically tuned by electric field alteration
of the density of DWs in the material. However, the effect
only existed at low temperatures, i.e., as long as phonon-phonon
interactions did not become the dominant scattering mechanism.
This drawback has been recently overcome with the advent of
nanostructured ferroelectrics. Ihlefeld \etal\ have been able to
decrease the DW spacing in ferroelectric thin-films made of
BiFeO$_3$~\cite{Hopkins2013} and Pb(Zr$_0.3$Ti$_0.3$)O$_3$~\cite{Ihlefeld2015} 
below the average phonon mean free path, which has made it possible to expand 
the electrically actuated thermal switch operation over a broad 
temperature range, including room temperature, thus boosting its 
potential technological impact.

In spite of the above experimental evidence, a quantitative evaluation 
of the DWs TBR is missing. Alternatively, this relevant magnitude can be
estimated by means of atomistic computational simulations of phonon 
transport. Nevertheless, for the case of ferroelectric DWs, reports 
of such simulations are scarce.~\cite{Wang2016,Royo2017} This
can be attributed to the fact that large supercells are required in this type of 
calculations, which hampers the use of accurate \emph{ab initio} 
methods, such as density-functional theory (DFT); a second problem relates to the relatively 
poor transferability and accuracy of the interatomic potential models 
available nowadays for perovskite oxides. Some of us have recently 
reported the first study of phonon transport through ferroelectric DWs
with atomistic precision.~\cite{Royo2017} To this end, we obtained the 
atomic force constants from second-principles model potentials~\cite{WojdeJPCM13} 
and performed harmonic (ballistic) phonon transport simulations within the 
nonequilibrium Green's functions formalism. The calculations revealed 
an unprecedented polarization-dependent phonon scattering mechanism 
occurring at PbTiO$_3$ 180\degree DWs capable to longitudinally polarize
a thermal flux when piercing several DWs. Yet, the harmonic description 
employed in that study limits, in principle, the validity of the results 
obtained to the low-temperature and short-channel regimes in which 
phonon-phonon scattering events can be neglected.
In the present paper we go beyond this limitation by performing 
molecular dynamics simulations -- so that we include all orders of 
anharmonicity in the description of the lattice dynamics -- devoted 
to investigate phonon transport across ferroelectric DWs in the 
technologically relevant diffusive regime.

\begin{figure}[t!]
\centering
\includegraphics[width=0.75\linewidth]{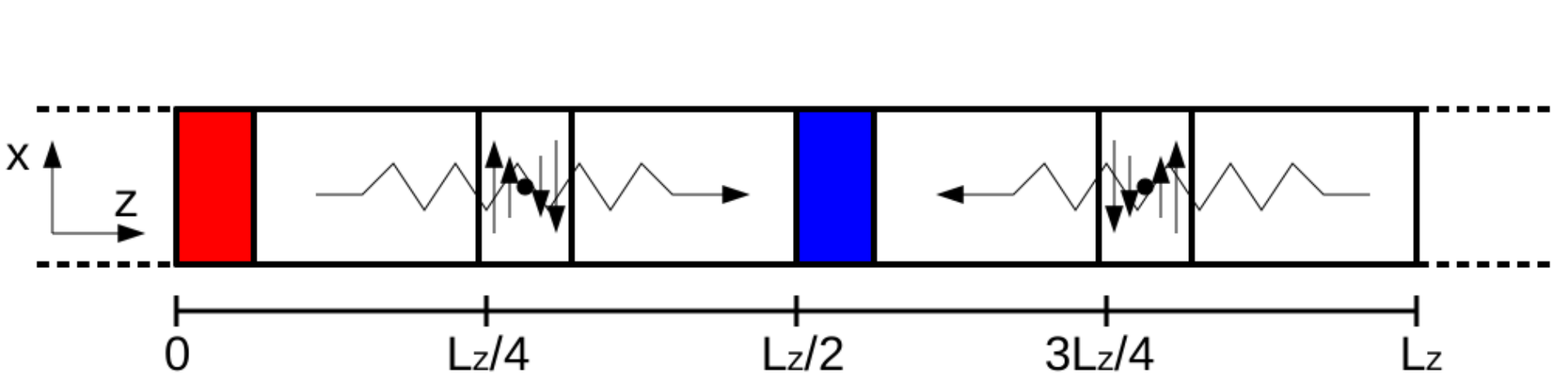}
\caption{Schematic view of the supercell for a simulation in which thermal
         transport through two independent DWs is evaluated. The red and blue rectangles
         illustrate the position of the thermal source and sink, respectively, and
         the transport directions are indicated with horizontal arrows. 
         }
\label{fig:setup}
\end{figure}

We use nonequilibrium molecular dynamics (NEMD) to study the thermal 
transport properties of PbTiO$_3$ in mono- and multidomain ferroelectric 
states at 200~K. To this end, as schematically shown in Fig.~\ref{fig:setup}, 
we generate a steady state heat flux along the $z$ axis by injecting a 
certain amount of kinetic energy in a heat source placed at $z=0$, 
which is then removed through a heat sink at $z=L_z/2$, where $L_z$ 
is the size of the simulation supercell along the transport 
direction~\cite{SchellingPRB02}. The resulting heat flux is 
calculated as

\begin{equation}
J = \frac{\Delta \epsilon}{2 A \Delta t},
\end{equation}

\noindent where $\Delta \epsilon$ is the energy injected/extracted, $A$ is
the supercell cross-section and $\Delta t$ is the timestep.
The thermal conductivity $\kappa$ is then computed from Fourier's 
law after estimating the thermal gradient that builds up in response
to the imposed heat flux
\begin{equation}
\kappa = J/\nabla T.
\label{eq:fourier}
\end{equation}

The energy and the forces, as well as the second order force-constants required for the harmonic
calculations presented below, of the PbTiO$_3$ lattice are calculated with the second-principles
model potential developed by some of us and thoroughly described in Ref.~\onlinecite{WojdeJPCM13}.
Within this model the dependence of the lattice energy on the atomic distortions
associated with ferroelectricity is expressed as a Taylor series around the
paraelectric cubic perovskite structure. More precisely, the energy is conveniently 
split in three contributions; phonon energy, strain energy and strain-phonon interaction
energy, and each one is Taylor expanded as a function of all possible atomic displacements
and strains. The series are truncated at 4th order and only pairwise interaction terms,
including long-range dipole-dipole interactions, are considered in the model. The potential parameters 
in the series were either directly determined from first-principles DFT calculation or 
others, the higher order ones, were fitted to reproduce a training set of relevant 
lattice-dynamical and structural data (see, e.g., Refs.~\onlinecite{WojdeJPCM13,EscorihuelaPRB17}).
The resulting model has repeatedly demonstrated its reliability and predictive power 
(tested against direct DFT simulations) in a number of works, most of which involved DWs.
~\cite{WojdeJPCM13,WojdePRL14,Zubko2016}

We use a $6 \times 6 \times 180$ supercell and a 0.1~fs timestep.
Initially we thermalize the system 
at 200~K by rescaling the velocities of the atoms for 1~ps. When the system
is equilibrated, we start the NEMD run by injecting/extracting
a certain amount of kinetic energy in the heat source/sink, whose
size amounts to 5 unit cells along $z$. Outside the heat source and
sink the system evolves microcanonically. We run 100~ps to reach
a nonequilibrium steady-state, with $J_z=6.3 \cdot 10^{10}$~W~m$^{-2}$,
and then average the temperature over the next 200~ps.
The strain, previously determined with a standard equilibrium Monte 
Carlo calculation at effective temperature of 200~K, is kept fixed 
in the NEMD simulation. This in principle avoids the need of carrying 
out the NEMD run in the NPT ensemble.

\begin{figure}[t!]
\centering
\includegraphics[width=0.75\linewidth]{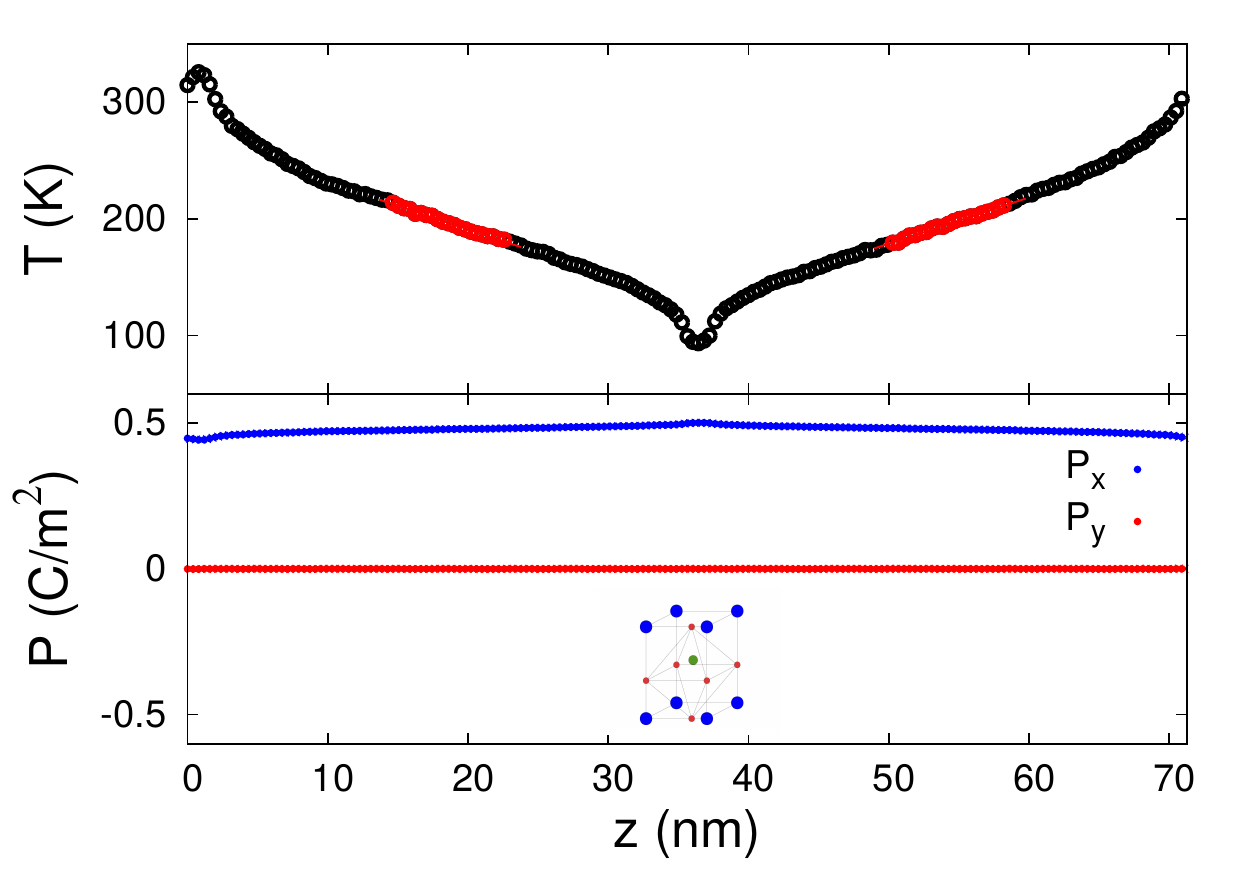}
\caption{(Top) Temperature and (bottom) polarization profiles along the transport direction $z$
         for the case of a monodomain PbTiO$_3$. Heat flows from the heat source at
         $z=0$ to the heat sink at $z=35$~nm. Data in red are 
         used to fit the thermal gradient. The uniform ferroelectric
         distortion is sketched in the inset of the bottom panel.
         }
\label{fig:TP_profile_mono}
\end{figure}

Our reference system is a PbTiO$_3$ supercell in its ferroelectric
ground-state, i.e., a monodomain configuration with all the ferroelectric distortions pointing along $x$
and, thereby, a continuous polarization $P_x$ developed throughout the simulation box. 
Far enough from the heat source and sink, the averaged temperature 
profile is linear as predicted by Fourier's law, with a fitted
slope of 4.1~K/nm (see Fig.~\ref{fig:TP_profile_mono}). This estimate, 
together with the imposed heat flux, allows us to compute the thermal 
conductivity from Eq.~\ref{eq:fourier}. In this way, we obtain 
$\kappa_{\text{PTO}}=$16~W~m$^{-1}$K$^{-1}$, in good agreement with the self-consistent 
solution of the Boltzmann transport equation~\cite{LiCPC14}, using 2nd 
and 3rd order interatomic force constants calculated within the same 
second-principles model as inputs. Tachibana \etal~\cite{TachibanaAPL08}
experimentally reported a lower value of around 6~W~m$^{-1}$K$^{-1}$.
However, they report the existence of complex domain structures in 
their PbTiO$_3$ samples, thus the true thermal conductivity of the
monodomain could be larger. 
On the other hand, finite size effects, which in general bedevil NEMD 
simulations~\cite{SchellingPRB02, SellanPRB10}, have a negligible 
effect in our study. We are mostly interested in the TBR, which 
determines the ratio between the thermal resistance of the high and 
low conduction states, and has thus a pivotal role in the operation of 
a potential thermal switch. As we will conclusively prove below, the TBR of 
the DW here considered is a strictly local property that basically 
depends on harmonic interactions which have been shown to be properly 
reproduced by the second-principles model potential.~\cite{WojdePRL14}

\begin{figure}[t!]
\centering
\includegraphics[width=0.75\linewidth]{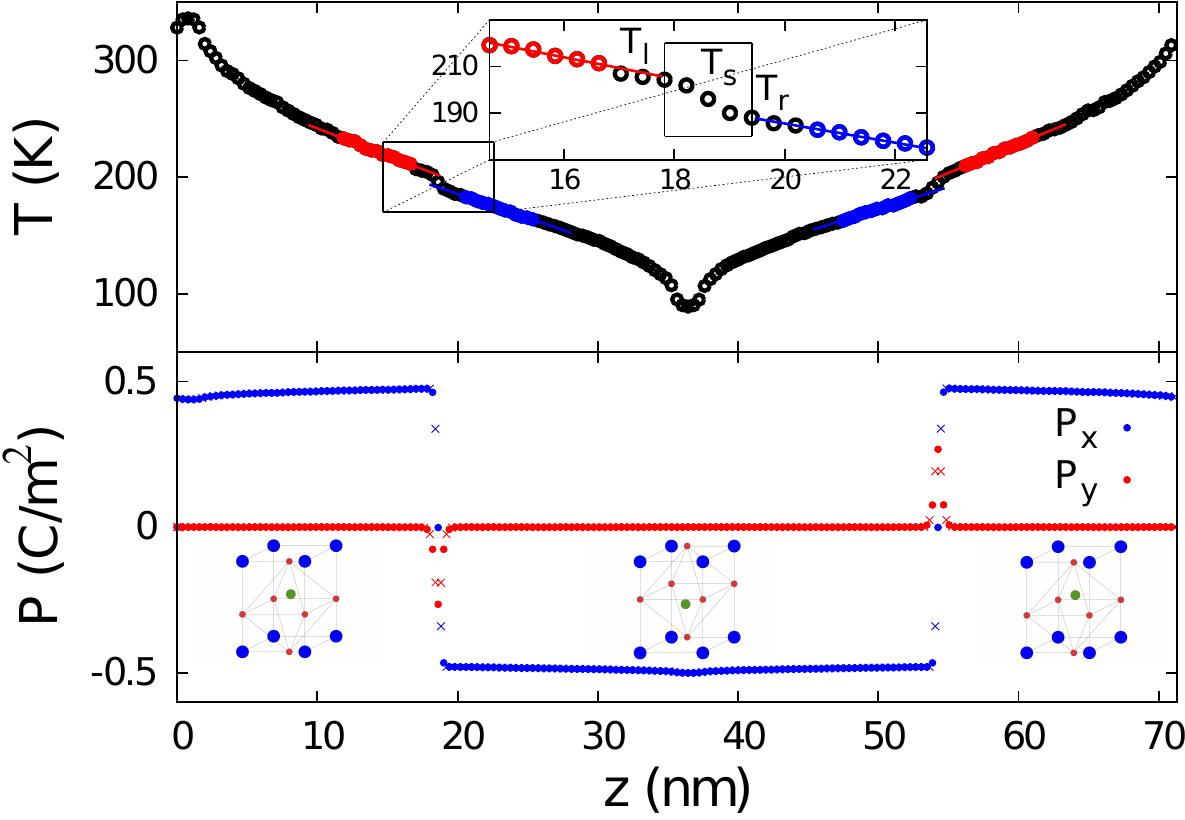}
\vskip 1cm
\caption{(Top) Temperature and (bottom) polarization profiles along the transport direction $z$
         for the case with one DW. Heat source and sink are placed like in
         Fig.~\ref{fig:TP_profile_mono} and the DWs are at $\sim$~17 and 
         $\sim$~52~nm. The inset of the top panel shows a zoomed view
         of the temperature profile at one of the DW; the relevant
         temperatures used to calculated the TBR are indicated.
         }
\label{fig:TP_profile_DW}
\end{figure}

We now move to the case of an individual 180\degree DW. Notice that, 
to satisfy periodic boundary conditions, an even number of DWs must 
be present in the computational cell. Therefore, each calculation will
provide two values of the TBR, assessing the error bar of our
estimate. As schematically indicated in Fig.~\ref{fig:setup}, the 
DWs are placed at $z=L_z/4$ and $z=3L_z/4$.
The temperature profile, shown in Fig.~\ref{fig:TP_profile_DW}, 
features a jump at each DW, the characteristic signature of TBR. 
A ferroelectric DW is the paradigm of a structurally sharp interface and apparently has
a vanishing thickness. The polarization and the temperature change 
rather abruptly, indeed, but they do so within a finite number of layers of material,
as shown in the zoomed view of the inset of Fig.~\ref{fig:TP_profile_DW}.
For this reason, we calculate the TBR within a generalized form of
the more common Kapitza resistance formalism.~\cite{RuraliPCCP16} We proceed as follows.
First, we estimate the interface thickness $\Delta z_{\text{DW}}$ by tracking the 
spontaneous polarization appearing at the DW\cite{WojdePRL14} and 
perpendicular to the direction of the 
ferroelectric distortion, $P_y$;~\cite{px} a change of more than two standard 
deviations from the reference values (far from the DW and the heat 
source and sink) identify the interface. We obtain an effective 
thickness of 20~\AA, i.e. 5 unit cells. Next we evaluate the
DW temperature, $T_{\text{s}}$, and the temperatures
at the DW boundaries, $T_{\text{l}}$ and $T_{\text{r}}$ (see inset
of Fig.~\ref{fig:TP_profile_DW}).

The TBR is then computed within the nonequilibrium thermodynamics 
formalism~\cite{SigneDick,DettoriAiPX16,RuraliPCCP16} by first writing 
the entropy production due to heat flow at the interface as
\begin{equation}
\sigma ^{\text{s}}=J^{\text{i}}\left( \frac{1}{T_{\text{s}}}-\frac{1}{T_{\text{l}}}\right) +J^{\text{o}}\left( \frac{1}{T_{\text{r}}}-\frac{1}{T_{\text{s}}}\right)
\end{equation}
where $J^{\text{i}}$ ($J^{\text{o}}$) is the heat flux entering (exiting)
the interface. In the stationary state $J^{\text{i}}=J^{\text{o}}=J$ and 
the corresponding force-flux relations are
\begin{equation}
\frac{1}{T_{\text{s}}}-\frac{1}{T_{\text{l}}} =r^{\text{s,i}}J
\label{eq:forceflux1}
\end{equation}
\begin{equation}
\frac{1}{T_{\text{r}}}-\frac{1}{T_{\text{s}}} =r^{\text{s,o}}J
\label{eq:forceflux2}
\end{equation}
where, according to the formalism of nonequilibrium thermodynamics, the
thermal driving force is the inverse temperature.
The TBR is then written as the sum of $r^{\text{s,i}}$ and $r^{\text{s,o}}$
and reads
\begin{equation}
R_{\text{DW}}=\frac{1}{J}\left( \frac{1}{T_{\text{r}}}-\frac{1}{T_{\text{l}}}\right) T_{\text{s}}^2 = R_{\text{s}} T_{\text{s}}^2
\label{eq:ri}
\end{equation}
where $R_{\text{s}}$ is the Onsager coefficient for thermal conductivity
and the factor $T_{\text{s}}^2$ is added to recover the dimension
of the Kapitza resistance.
We obtain a value for $R_{\text{DW}}$ of 2.9$\cdot 10^{-10}$~K~m$^2$/W.

The TBR adds to the intrinsic thermal resistance of the monodomain,
yielding a larger total thermal resistance. Therefore, a DW results
in a low conductive state and can be used as the '0' of a phononic binary
code. Erasing the DW switches the phononic bit to '1'.
The ratio between the high and the low conductive states is obtained
dividing the resistance of a segment of material $\bar{L}$ 
with and without the TBR and reads

\begin{equation}
\frac{R^{high}}{R^{low}} = 1 + \frac{R_{\text{DW}}}{\bar{L} / \kappa_{\text{PTO}}}
\label{eq:ratio}
\end{equation}

\noindent
In the setup of Fig.~\ref{fig:setup}, $\bar{L}$ can be as large as
$L_z/4-\Delta z$, but we take the lower value of $\bar{L}=15.2$~nm to
avoid entering the temperature non-linear region next to the heat
source and sink (see Fig.~\ref{fig:TP_profile_mono}). With this 
choice we obtain a ratio  
$R^{high}/R^{low} \sim 1.18$.
% Notice that $R^{high}-R^{low}$, which
% amounts to 2.3$\cdot 10^{-10}$~K~m$^2$/W, provides another estimate 
% of the TBR. The good agreement with the value calculated above
% proves a good degree of additivity of the resistances involved.

We have also calculated the TBR within the nonequilibrium Green's
function (NEGF) approach.\cite{PourfathBOOK14,SadasivamARHT14} This scheme is based on the harmonic
approximation, hence no phonon-phonon scattering is accounted for
and only elastic scattering mechanisms -- e.g. associated to impurities or boundaries -- are described. 
This means that a homogeneous perfect crystal,
within this approximation, has an infinite thermal conductivity
and a length-independent thermal conductance. Following the NEGF approach we
partition the system in three regions: two homogeneous (i.e., DW free) semi-infinite contacts acting as
coherent phonon reservoirs and a central scattering region wherein different number of DWs
are included. We calculate the thermal conductance due to phonons traveling between the two
contacts across the scattering region with the Landauer formula,

\begin{equation}
G(T) = \frac{\hbar}{2\pi\,\Omega} \int \omega \mathcal{T}(\omega)
\left(\frac{\partial n_0(\omega,T)}{\partial T} \right) d\omega.
\label{eq:land}
\end{equation}

\noindent
Here, $n_0$ is the equilibrium Bose-Einstein distribution, $\Omega$ 
is the channel cross section, and
$\hbar$ is the reduced Planck constant. Since we assume periodic
boundary conditions in the directions perpendicular to the thermal 
flux, the total phonon transmission function is computed as

\begin{equation}
\mathcal{T}(\omega)=\frac{\Omega}{(2\pi)^2}\int \Xi(\omega,\mathbf{k}_{\perp}) \,
d\mathbf{k}_{\perp},
 \label{eq:totaltrans}
\end{equation}

\noindent with $\Xi(\omega,\mathbf{k}_{\perp})$ being the phonon transmission function
calculated at a discrete point ($\mathbf{k}_{\perp}$) of the 2D transverse Brillouin zone
by means of the Caroli formula,

\begin{equation}
 \Xi(\omega,\mathbf{k}_{\perp})
 = Tr \left[  \boldsymbol{\Gamma}_L(\omega,\mathbf{k}_{\perp}) \, \mathbf{G}_C^r(\omega,\mathbf{k}_{\perp}) \,
 \boldsymbol{\Gamma}_R(\omega,\mathbf{k}_{\perp})\, \mathbf{G}_C^a(\omega,\mathbf{k}_{\perp})  \right].
\label{eq:trans}
\end{equation}

\noindent
Here, $\mathbf{G}_C^{r(a)}$ is the retarded (advanced) Green's function of the scattering 
region which is calculated as

\begin{equation}
 \mathbf{G}_C^{r(a)}(\omega,\mathbf{k_{\perp}})= \left[ \omega^2\,\mathbf{I} - \mathbf{H}_C(\mathbf{k_{\perp}}) - 
 \mathbf{\Sigma}_L^{r(a)}(\omega,\mathbf{k_{\perp}}) -  \mathbf{\Sigma}_R^{r(a)}(\omega,\mathbf{k_{\perp}})  \right]^{-1},
\label{eq:GF}
\end{equation}

\noindent with $\mathbf{I}$ being the identity matrix, $\mathbf{H}_C$ the second-order force-constant matrix 
of the atoms in the scattering region, and $\mathbf{\Sigma}_{L(R)}^{r(a)}$ the left (right) contact self-energy. 
Besides, in Eq.~\ref{eq:trans}, $\boldsymbol{\Gamma}_{L(R)}$ are broadening functions defined as 
$\boldsymbol{\Gamma}_{L(R)}(\omega,\mathbf{k_{\perp}})=i \left( \mathbf{\Sigma}_{L(R)}^r(\omega,\mathbf{k_{\perp}}) - \mathbf{\Sigma}_{L(R)}^a(\omega,\mathbf{k_{\perp}}) \right)$. 

Finally, the self-energies accounting for the coupling between the contacts and the scattering region are calculated as

\begin{subequations}
\label{eq:selfe}
 \begin{align}
  \boldsymbol{\Sigma}_L^{r(a)}(\omega,\mathbf{k_{\perp}})= \mathbf{H}_{C,L}(\mathbf{k_{\perp}}) \, 
  \mathbf{g}_L^{r(a)}(\omega,\mathbf{k_{\perp}}) \, \mathbf{H}_{L,C}(\mathbf{k_{\perp}}) \\
  \boldsymbol{\Sigma}_R^{r(a)}(\omega,\mathbf{k_{\perp}})= \mathbf{H}_{C,R}(\mathbf{k_{\perp}}) \, 
  \mathbf{g}_R^{r(a)}(\omega,\mathbf{k_{\perp}}) \, \mathbf{H}_{R,D}(\mathbf{k_{\perp}}) ,
 \end{align}
\end{subequations}

\noindent where $\mathbf{H}_{\alpha,\beta}$ are force-constant matrices that describe the interaction between the atoms
in the contacts and those in the scattering region, and $\mathbf{g}_{L(R)}^{r(a)}$ are the surface contacts Green's functions.
The latter are iteratively calculated following the Sancho-Rubio approach.~\cite{SanchoJPF84}

\begin{figure}[t!]
\centering
\includegraphics[width=0.75\linewidth]{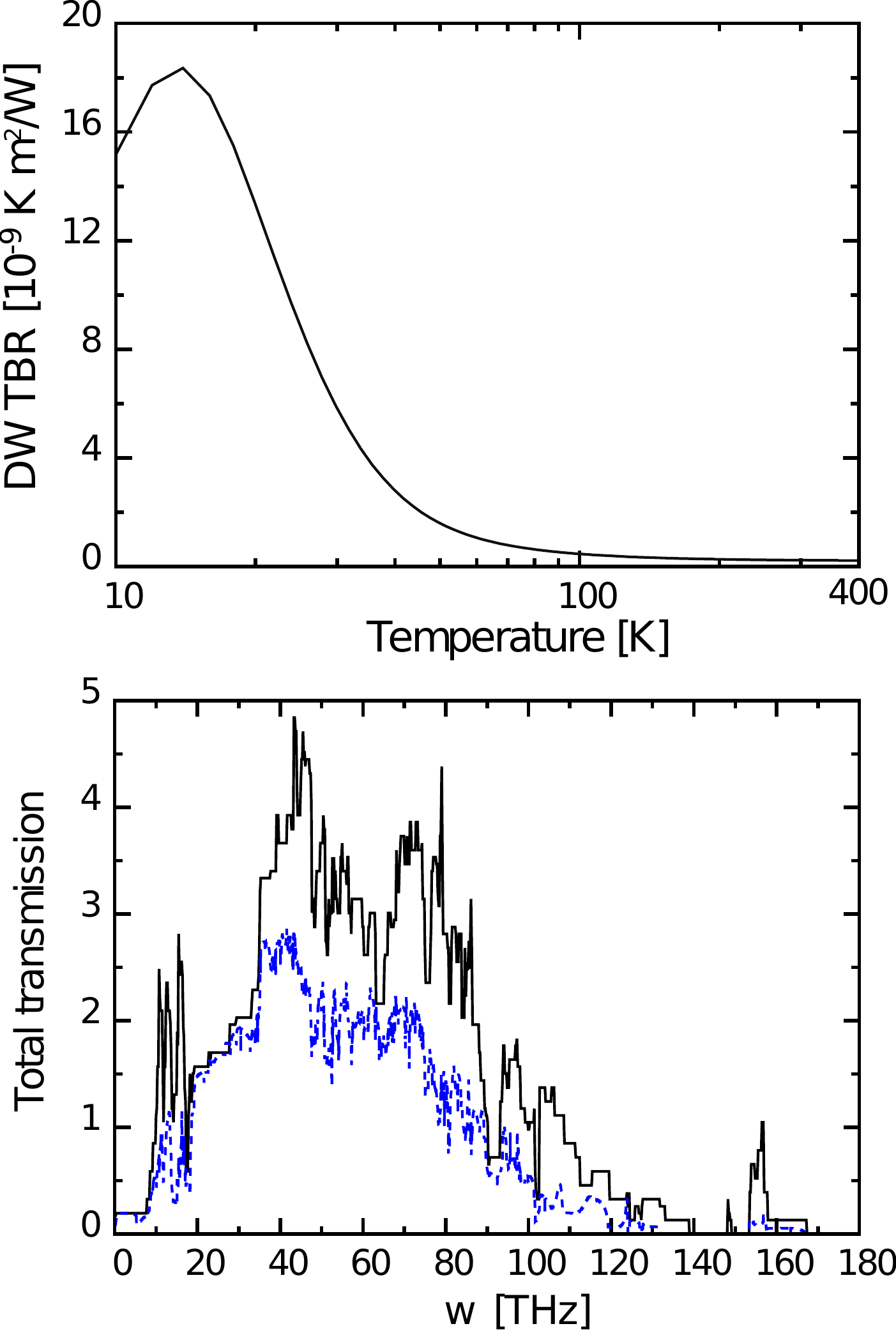}
\caption{(Top) TBR as a function of temperature computed within the 
	 harmonic NEGF approach as, $G_{DW}^{-1} - G_{mono}^{-1}$, where 
         $G_{DW}$ ($G_{mono}$) is the thermal conductance of the case with (without) DW. 
         (Bottom) Phonon transmission as a function of frequency for the case of a 
	 monodomain (continuous line) and a DW (dashed line). 
         }
\label{fig:negf}
\end{figure}

The total phonon transmission functions (Eq.~\ref{eq:totaltrans}) for a monodomain and for an individual DW
systems are shown in Fig.~\ref{fig:negf}(b). It is observed that the presence of the DW largely suppresses 
the transmission of phonons with frequencies higher than $\sim 40$ THz, as would be expected for abrupt interfaces 
such as the present DW. Nevertheless, a strong scattering is also observed for phonons with frequencies
between 10 and 20 THz meaning that the DW does not completely act as a low-pass filter.  

Elastic scattering at the DW is the only scattering mechanism 
captured at this level of the theory, thus the TBR
is simply the difference between the thermal resistance
of a system with and without a DW. As can be observed in Fig.~\ref{fig:negf}, at
temperatures higher than 100~K we obtain an asymptotic TBR value of 2.8$\cdot 
10^{-10}$~K~m$^2$/W, practically the same value obtained with NEMD at 200~K.
The excellent agreement between the length-independent estimate of 
NEGF and the NEMD simulations, where anharmonic effects are included, implies 
that (i)~the TBR is a local property of the DW; (ii)~anharmonic 
effects play a negligible role in the DW TBR. The DW resistance 
increases at low temperatures ($<60$ K) due to the above mentioned large scattering
experienced by the low-energy modes that are excited in this
regime.

\begin{figure}[t!]
\centering
\includegraphics[width=0.75\linewidth]{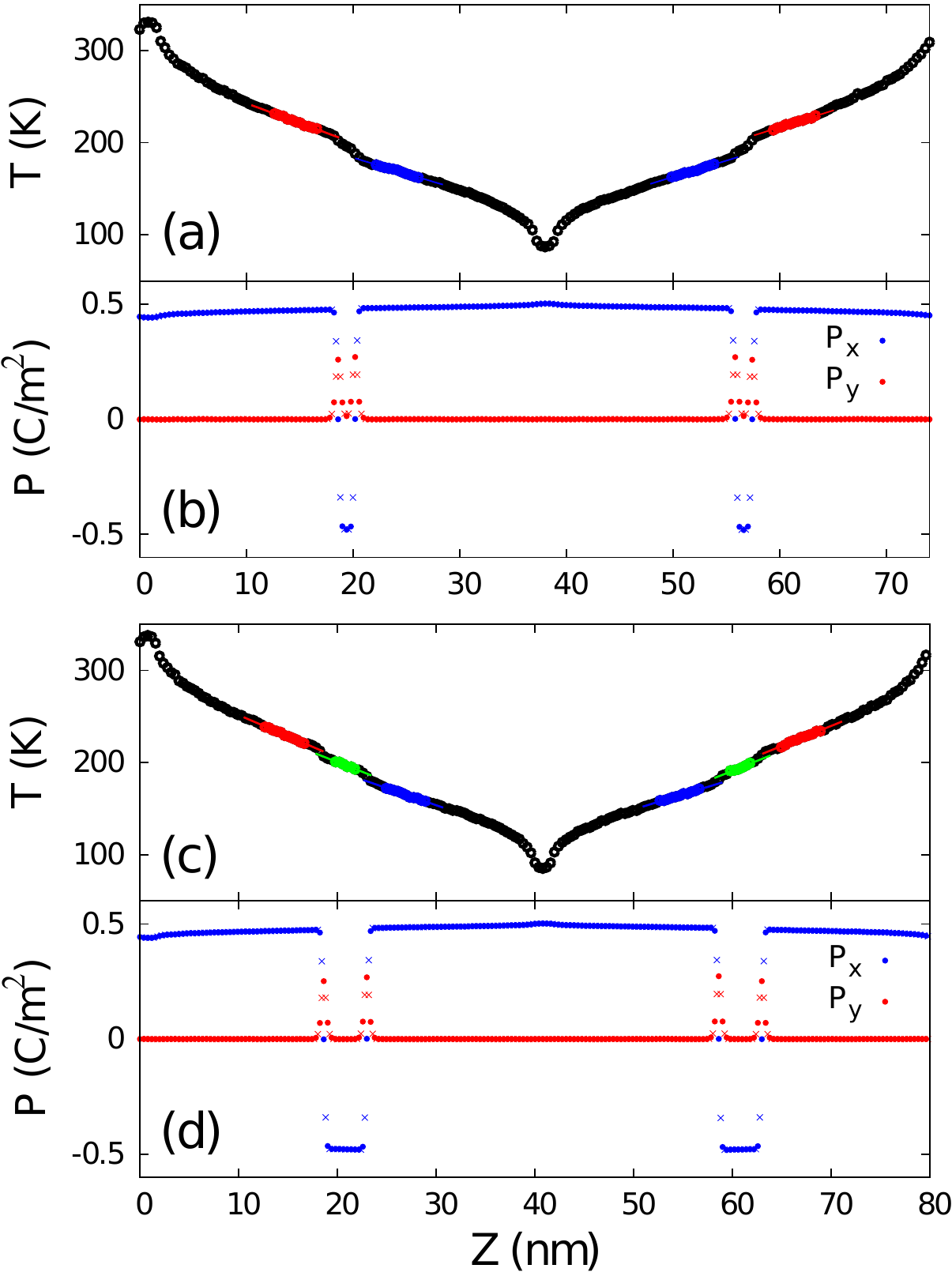}
\caption{ Temperature and polarization profiles along the transport direction $z$ for the cases with two DWs
          separated by $\sim 1.5$~nm, (a) and (b), and $\sim 4$~nm, (c) and (d).
         }
\label{fig:Fig5}
\end{figure}

The possibility of introducing more DWs is very appealing, because 
it would allow increasing the ratio $R^{high}/R^{low}$. Therefore,
it is interesting to asses to what extent subsequent DWs behave like
independent scattering centers with resistances that barely sum up,
a sound assumption in a purely diffusive transport regime.~\cite{LuAIP15}
For this reason, we next study the thermal conduction in systems with
two DWs. In particular, we consider two cases of DW pairs: in one case 
the spacing between them is 1.5~nm, in the other 4~nm. We have repeated the procedure
described above to define the interface thickness and to calculate 
the TBR; the final steady state temperature profiles are shown in Fig.~\ref{fig:Fig5}. 
In the case of the smaller separation, the DWs \emph{coalesce} making  
difficult to distinguish two separate interface regions; they are thus treated as one single
interface complex. Proceeding in this way we obtain a TBR of 4.7$\cdot 10^{-10}$~K~m$^2$/W.
When the spacing between the DWs is larger we can treat them individually and we obtain a TBR of
2.3$\cdot 10^{-10}$~K~m$^2$/W for each one. In both cases we did not appreciate any 
difference between parallel or antiparallel orientation of the spontaneous polarization 
$P_y$ occurring at the DW.~\cite{WojdePRL14} 

While roughly speaking the thermal resistances of consecutive DWs add 
up, the obtained TBR per DW is smaller than the value of the individual 
DW previously calculated.
In the low temperature, ballistic transport regime, these DWs have
been shown to behave as phonon filters~\cite{Royo2017} and the presently 
observed behavior suggest that this effect might partially persist 
at 200~K and for the specific DW separation here considered.
Actually, the total TBR for the system with two separated DWs via NEMD simulations
(4.6$\cdot 10^{-10}$~K~m$^2$/W) is larger than the one obtained from the NEGF harmonic
calculations (3.7$\cdot 10^{-10}$~K~m$^2$/W).
These trends in the additivity of the TBRs indicate that the temperatures 
and dimensions assumed in our NEMD simulations 
do not entail a completely diffusive transport, although we can appreciate deviations from the 
harmonic transport regime.

We have calculated the $R^{high}/R^{low}$ for the configuration with two DWs 
like in Eq.~\ref{eq:ratio} and taking the same value for $\bar{L}$;
the contribution from the TBR is now twice 2.3$\cdot 10^{-10}$~K~m$^2$/W, 
while for $\kappa_{PTO}$, as in the above discussion for a single DW, we have 
taken the value obtained in the monodomain configuration. 
With these assumptions we obtain a value of 1.26.
Therefore, by adding more DWs the $R^{high}/R^{low}$ ratio can indeed become 
larger, but the increase in the case here considered is moderate. The 
gain $R^{high}/R^{low}$ depends on a delicate balance between design 
parameters of the system: well spaced DWs entail a large total $R_{\text{DW}}$
(approximately $n$ times the TBR of an individual DW, being $n$ the number
of DWs), but also a large
$\bar{L}$; conversely, in a series of nearby DWs $\bar{L}$ can be small,
but the TBR per DW could also decrease because of constructive interference
and filtering effect.

To summarize, in this paper we have numerically tested the extent to which the simplest
type of ferroelectric DWs occurring in PbTiO$_3$, namely, 180\degree DWs, behave as barriers
for lattice thermal conduction. Our molecular dynamics simulations carried out at relevant temperatures 
for device operation have shown that the thermal resistance increases by a factor  
of around 20\% due to the presence of a single DW, at the system sizes here considered. 
This factor can be further increased by incorporating more DWs in the system, though the 
gain is lower than expected due to the partially non-diffusive nature of thermal transport in the regions between the DWs.
We have also demonstrated the local and harmonic character of
the DW scattering, as evidenced by the excellent agreement observed among the TBRs from anharmonic NEMD and
harmonic NEGF calculations. Our numerical results support the use of 
ferroelectric domain walls as active mobile elements in electrically actuated thermal switches.

\begin{acknowledgments}
We acknowledge financial support by the Ministerio de Econom\'ia y 
Competitividad (MINECO) under grant FEDER-MAT2013-40581-P and the 
Severo Ochoa Centres of Excellence Program under Grant SEV-2015-0496
and by the Generalitat de Catalunya under grants no. 2014 SGR 301.
M.R. acknowledges financial support from the Beatriu de Pin\'os 
fellowship program (2014 BP\_B 00101).
C.E.S and J.I. are funded by the Luxembourg National Research Fund
through the CORE (Grant C15/MS/10458889 NEWALLS), PEARL (Grant
P12/4853155 COFERMAT) and AFR (PhD Grant No. 9934186 for C.E.S.) programs.
\end{acknowledgments}

\bibliography{complete}

\begin{thebibliography}{31}%
\makeatletter
\providecommand \@ifxundefined [1]{%
 \@ifx{#1\undefined}
}%
\providecommand \@ifnum [1]{%
 \ifnum #1\expandafter \@firstoftwo
 \else \expandafter \@secondoftwo
 \fi
}%
\providecommand \@ifx [1]{%
 \ifx #1\expandafter \@firstoftwo
 \else \expandafter \@secondoftwo
 \fi
}%
\providecommand \natexlab [1]{#1}%
\providecommand \enquote  [1]{``#1''}%
\providecommand \bibnamefont  [1]{#1}%
\providecommand \bibfnamefont [1]{#1}%
\providecommand \citenamefont [1]{#1}%
\providecommand \href@noop [0]{\@secondoftwo}%
\providecommand \href [0]{\begingroup \@sanitize@url \@href}%
\providecommand \@href[1]{\@@startlink{#1}\@@href}%
\providecommand \@@href[1]{\endgroup#1\@@endlink}%
\providecommand \@sanitize@url [0]{\catcode `\\12\catcode `\$12\catcode
  `\&12\catcode `\#12\catcode `\^12\catcode `\_12\catcode `\%12\relax}%
\providecommand \@@startlink[1]{}%
\providecommand \@@endlink[0]{}%
\providecommand \url  [0]{\begingroup\@sanitize@url \@url }%
\providecommand \@url [1]{\endgroup\@href {#1}{\urlprefix }}%
\providecommand \urlprefix  [0]{URL }%
\providecommand \Eprint [0]{\href }%
\providecommand \doibase [0]{http://dx.doi.org/}%
\providecommand \selectlanguage [0]{\@gobble}%
\providecommand \bibinfo  [0]{\@secondoftwo}%
\providecommand \bibfield  [0]{\@secondoftwo}%
\providecommand \translation [1]{[#1]}%
\providecommand \BibitemOpen [0]{}%
\providecommand \bibitemStop [0]{}%
\providecommand \bibitemNoStop [0]{.\EOS\space}%
\providecommand \EOS [0]{\spacefactor3000\relax}%
\providecommand \BibitemShut  [1]{\csname bibitem#1\endcsname}%
\let\auto@bib@innerbib\@empty
%</preamble>
\bibitem [{\citenamefont {Li}\ \emph {et~al.}(2012)\citenamefont {Li},
  \citenamefont {Ren}, \citenamefont {Wang}, \citenamefont {Zhang},
  \citenamefont {H{\"a}nggi},\ and\ \citenamefont {Li}}]{LiRMP12}%
  \BibitemOpen
  \bibfield  {author} {\bibinfo {author} {\bibfnamefont {N.}~\bibnamefont
  {Li}}, \bibinfo {author} {\bibfnamefont {J.}~\bibnamefont {Ren}}, \bibinfo
  {author} {\bibfnamefont {L.}~\bibnamefont {Wang}}, \bibinfo {author}
  {\bibfnamefont {G.}~\bibnamefont {Zhang}}, \bibinfo {author} {\bibfnamefont
  {P.}~\bibnamefont {H{\"a}nggi}}, \ and\ \bibinfo {author} {\bibfnamefont
  {B.}~\bibnamefont {Li}},\ }\href@noop {} {\bibfield  {journal} {\bibinfo
  {journal} {Rev. Mod. Phys.}\ }\textbf {\bibinfo {volume} {84}},\ \bibinfo
  {pages} {1045} (\bibinfo {year} {2012})}\BibitemShut {NoStop}%
\bibitem [{\citenamefont {Dames}(2009)}]{DamesJHT09}%
  \BibitemOpen
  \bibfield  {author} {\bibinfo {author} {\bibfnamefont {C.}~\bibnamefont
  {Dames}},\ }\href@noop {} {\bibfield  {journal} {\bibinfo  {journal} {J. Heat
  Transfer}\ }\textbf {\bibinfo {volume} {131}},\ \bibinfo {pages} {061301}
  (\bibinfo {year} {2009})}\BibitemShut {NoStop}%
\bibitem [{\citenamefont {Rurali}\ \emph {et~al.}(2014)\citenamefont {Rurali},
  \citenamefont {Cartoix\`a},\ and\ \citenamefont {Colombo}}]{RuraliPRB14}%
  \BibitemOpen
  \bibfield  {author} {\bibinfo {author} {\bibfnamefont {R.}~\bibnamefont
  {Rurali}}, \bibinfo {author} {\bibfnamefont {X.}~\bibnamefont {Cartoix\`a}},
  \ and\ \bibinfo {author} {\bibfnamefont {L.}~\bibnamefont {Colombo}},\
  }\href@noop {} {\bibfield  {journal} {\bibinfo  {journal} {Phys. Rev. B}\
  }\textbf {\bibinfo {volume} {90}},\ \bibinfo {pages} {041408} (\bibinfo
  {year} {2014})}\BibitemShut {NoStop}%
\bibitem [{\citenamefont {Yang}\ \emph {et~al.}(2008)\citenamefont {Yang},
  \citenamefont {Zhang},\ and\ \citenamefont {Li}}]{YangAPL08}%
  \BibitemOpen
  \bibfield  {author} {\bibinfo {author} {\bibfnamefont {N.}~\bibnamefont
  {Yang}}, \bibinfo {author} {\bibfnamefont {G.}~\bibnamefont {Zhang}}, \ and\
  \bibinfo {author} {\bibfnamefont {B.}~\bibnamefont {Li}},\ }\href@noop {}
  {\bibfield  {journal} {\bibinfo  {journal} {Appl. Phys. Lett.}\ }\textbf
  {\bibinfo {volume} {93}},\ \bibinfo {pages} {243111} (\bibinfo {year}
  {2008})}\BibitemShut {NoStop}%
\bibitem [{\citenamefont {Cartoix\`a}\ \emph {et~al.}(2015)\citenamefont
  {Cartoix\`a}, \citenamefont {Colombo},\ and\ \citenamefont
  {Rurali}}]{CartoixaNL15}%
  \BibitemOpen
  \bibfield  {author} {\bibinfo {author} {\bibfnamefont {X.}~\bibnamefont
  {Cartoix\`a}}, \bibinfo {author} {\bibfnamefont {L.}~\bibnamefont {Colombo}},
  \ and\ \bibinfo {author} {\bibfnamefont {R.}~\bibnamefont {Rurali}},\
  }\href@noop {} {\bibfield  {journal} {\bibinfo  {journal} {Nano Lett.}\
  }\textbf {\bibinfo {volume} {15}},\ \bibinfo {pages} {8255} (\bibinfo {year}
  {2015})}\BibitemShut {NoStop}%
\bibitem [{\citenamefont {Terraneo}\ \emph {et~al.}(2002)\citenamefont
  {Terraneo}, \citenamefont {Peyrard},\ and\ \citenamefont
  {Casati}}]{TerraneoPRL02}%
  \BibitemOpen
  \bibfield  {author} {\bibinfo {author} {\bibfnamefont {M.}~\bibnamefont
  {Terraneo}}, \bibinfo {author} {\bibfnamefont {M.}~\bibnamefont {Peyrard}}, \
  and\ \bibinfo {author} {\bibfnamefont {G.}~\bibnamefont {Casati}},\
  }\href@noop {} {\bibfield  {journal} {\bibinfo  {journal} {Phys. Rev. Lett.}\
  }\textbf {\bibinfo {volume} {88}},\ \bibinfo {pages} {094302} (\bibinfo
  {year} {2002})}\BibitemShut {NoStop}%
\bibitem [{\citenamefont {Li}\ \emph {et~al.}(2004)\citenamefont {Li},
  \citenamefont {Wang},\ and\ \citenamefont {Casati}}]{LiPRL04}%
  \BibitemOpen
  \bibfield  {author} {\bibinfo {author} {\bibfnamefont {B.}~\bibnamefont
  {Li}}, \bibinfo {author} {\bibfnamefont {L.}~\bibnamefont {Wang}}, \ and\
  \bibinfo {author} {\bibfnamefont {G.}~\bibnamefont {Casati}},\ }\href@noop {}
  {\bibfield  {journal} {\bibinfo  {journal} {Phys. Rev. Lett.}\ }\textbf
  {\bibinfo {volume} {93}},\ \bibinfo {pages} {184301} (\bibinfo {year}
  {2004})}\BibitemShut {NoStop}%
\bibitem [{\citenamefont {Roberts}\ and\ \citenamefont
  {Walker}(2011)}]{RobertsIJTS11}%
  \BibitemOpen
  \bibfield  {author} {\bibinfo {author} {\bibfnamefont {N.}~\bibnamefont
  {Roberts}}\ and\ \bibinfo {author} {\bibfnamefont {D.}~\bibnamefont
  {Walker}},\ }\href@noop {} {\bibfield  {journal} {\bibinfo  {journal} {Int.
  J. Therm. Sci.}\ }\textbf {\bibinfo {volume} {50}},\ \bibinfo {pages} {648}
  (\bibinfo {year} {2011})}\BibitemShut {NoStop}%
\bibitem [{\citenamefont {Mante}\ and\ \citenamefont
  {Volger}(1971)}]{Mante1971}%
  \BibitemOpen
  \bibfield  {author} {\bibinfo {author} {\bibfnamefont {A.}~\bibnamefont
  {Mante}}\ and\ \bibinfo {author} {\bibfnamefont {J.}~\bibnamefont {Volger}},\
  }\href {\doibase 10.1016/0031-8914(71)90164-9} {\bibfield  {journal}
  {\bibinfo  {journal} {Physica}\ }\textbf {\bibinfo {volume} {52}},\ \bibinfo
  {pages} {577} (\bibinfo {year} {1971})}\BibitemShut {NoStop}%
\bibitem [{\citenamefont {Weilert}\ \emph
  {et~al.}(1993{\natexlab{a}})\citenamefont {Weilert}, \citenamefont {Msall},
  \citenamefont {Wolfe},\ and\ \citenamefont {Anderson}}]{Weilert1993}%
  \BibitemOpen
  \bibfield  {author} {\bibinfo {author} {\bibfnamefont {M.~A.}\ \bibnamefont
  {Weilert}}, \bibinfo {author} {\bibfnamefont {M.~E.}\ \bibnamefont {Msall}},
  \bibinfo {author} {\bibfnamefont {J.~P.}\ \bibnamefont {Wolfe}}, \ and\
  \bibinfo {author} {\bibfnamefont {A.~C.}\ \bibnamefont {Anderson}},\ }\href
  {\doibase 10.1007/BF01315234} {\bibfield  {journal} {\bibinfo  {journal}
  {Zeitschrift f{\"u}r Phys. B Condens. Matter}\ }\textbf {\bibinfo {volume}
  {91}},\ \bibinfo {pages} {179} (\bibinfo {year}
  {1993}{\natexlab{a}})}\BibitemShut {NoStop}%
\bibitem [{\citenamefont {Weilert}\ \emph
  {et~al.}(1993{\natexlab{b}})\citenamefont {Weilert}, \citenamefont {Msall},
  \citenamefont {Anderson},\ and\ \citenamefont {Wolfe}}]{Weilert1993a}%
  \BibitemOpen
  \bibfield  {author} {\bibinfo {author} {\bibfnamefont {M.}~\bibnamefont
  {Weilert}}, \bibinfo {author} {\bibfnamefont {M.}~\bibnamefont {Msall}},
  \bibinfo {author} {\bibfnamefont {A.}~\bibnamefont {Anderson}}, \ and\
  \bibinfo {author} {\bibfnamefont {J.}~\bibnamefont {Wolfe}},\ }\href
  {\doibase 10.1103/PhysRevLett.71.735} {\bibfield  {journal} {\bibinfo
  {journal} {Phys. Rev. Lett.}\ }\textbf {\bibinfo {volume} {71}},\ \bibinfo
  {pages} {735} (\bibinfo {year} {1993}{\natexlab{b}})}\BibitemShut {NoStop}%
\bibitem [{\citenamefont {Hopkins}\ \emph {et~al.}(2013)\citenamefont
  {Hopkins}, \citenamefont {Adamo}, \citenamefont {Ye}, \citenamefont {Huey},
  \citenamefont {Lee}, \citenamefont {Schlom},\ and\ \citenamefont
  {Ihlefeld}}]{Hopkins2013}%
  \BibitemOpen
  \bibfield  {author} {\bibinfo {author} {\bibfnamefont {P.~E.}\ \bibnamefont
  {Hopkins}}, \bibinfo {author} {\bibfnamefont {C.}~\bibnamefont {Adamo}},
  \bibinfo {author} {\bibfnamefont {L.}~\bibnamefont {Ye}}, \bibinfo {author}
  {\bibfnamefont {B.~D.}\ \bibnamefont {Huey}}, \bibinfo {author}
  {\bibfnamefont {S.~R.}\ \bibnamefont {Lee}}, \bibinfo {author} {\bibfnamefont
  {D.~G.}\ \bibnamefont {Schlom}}, \ and\ \bibinfo {author} {\bibfnamefont
  {J.~F.}\ \bibnamefont {Ihlefeld}},\ }\href {\doibase 10.1063/1.4798497}
  {\bibfield  {journal} {\bibinfo  {journal} {Appl. Phys. Lett.}\ }\textbf
  {\bibinfo {volume} {102}},\ \bibinfo {pages} {121903} (\bibinfo {year}
  {2013})}\BibitemShut {NoStop}%
\bibitem [{\citenamefont {Ihlefeld}\ \emph {et~al.}(2015)\citenamefont
  {Ihlefeld}, \citenamefont {Foley}, \citenamefont {Scrymgeour}, \citenamefont
  {Michael}, \citenamefont {McKenzie}, \citenamefont {Medlin}, \citenamefont
  {Wallace}, \citenamefont {Trolier-McKinstry},\ and\ \citenamefont
  {Hopkins}}]{Ihlefeld2015}%
  \BibitemOpen
  \bibfield  {author} {\bibinfo {author} {\bibfnamefont {J.~F.}\ \bibnamefont
  {Ihlefeld}}, \bibinfo {author} {\bibfnamefont {B.~M.}\ \bibnamefont {Foley}},
  \bibinfo {author} {\bibfnamefont {D.~A.}\ \bibnamefont {Scrymgeour}},
  \bibinfo {author} {\bibfnamefont {J.~R.}\ \bibnamefont {Michael}}, \bibinfo
  {author} {\bibfnamefont {B.~B.}\ \bibnamefont {McKenzie}}, \bibinfo {author}
  {\bibfnamefont {D.~L.}\ \bibnamefont {Medlin}}, \bibinfo {author}
  {\bibfnamefont {M.}~\bibnamefont {Wallace}}, \bibinfo {author} {\bibfnamefont
  {S.}~\bibnamefont {Trolier-McKinstry}}, \ and\ \bibinfo {author}
  {\bibfnamefont {P.~E.}\ \bibnamefont {Hopkins}},\ }\href {\doibase
  10.1021/nl504505t} {\bibfield  {journal} {\bibinfo  {journal} {Nano Lett.}\
  }\textbf {\bibinfo {volume} {15}},\ \bibinfo {pages} {1791} (\bibinfo {year}
  {2015})}\BibitemShut {NoStop}%
\bibitem [{\citenamefont {Wang}\ \emph {et~al.}(2016)\citenamefont {Wang},
  \citenamefont {Wang}, \citenamefont {Ihlefeld}, \citenamefont {Hopkins},\
  and\ \citenamefont {Chen}}]{Wang2016}%
  \BibitemOpen
  \bibfield  {author} {\bibinfo {author} {\bibfnamefont {J.-J.}\ \bibnamefont
  {Wang}}, \bibinfo {author} {\bibfnamefont {Y.}~\bibnamefont {Wang}}, \bibinfo
  {author} {\bibfnamefont {J.~F.}\ \bibnamefont {Ihlefeld}}, \bibinfo {author}
  {\bibfnamefont {P.~E.}\ \bibnamefont {Hopkins}}, \ and\ \bibinfo {author}
  {\bibfnamefont {L.-Q.}\ \bibnamefont {Chen}},\ }\href {\doibase
  10.1016/j.actamat.2016.03.069} {\bibfield  {journal} {\bibinfo  {journal}
  {Acta Mater.}\ }\textbf {\bibinfo {volume} {111}},\ \bibinfo {pages} {220}
  (\bibinfo {year} {2016})}\BibitemShut {NoStop}%
\bibitem [{\citenamefont {Royo}\ \emph {et~al.}()\citenamefont {Royo},
  \citenamefont {Escorihuela-sayalero}, \citenamefont {{\'I}{\~n}iguez},\ and\
  \citenamefont {Rurali}}]{Royo2017}%
  \BibitemOpen
  \bibfield  {author} {\bibinfo {author} {\bibfnamefont {M.}~\bibnamefont
  {Royo}}, \bibinfo {author} {\bibfnamefont {C.}~\bibnamefont
  {Escorihuela-sayalero}}, \bibinfo {author} {\bibfnamefont {J.}~\bibnamefont
  {{\'I}{\~n}iguez}}, \ and\ \bibinfo {author} {\bibfnamefont {R.}~\bibnamefont
  {Rurali}},\ }\href@noop {} {\bibinfo  {journal} {Phys. Rev. Mater.,
  accepted}\ }\BibitemShut {NoStop}%
\bibitem [{\citenamefont {Wojde{\l}}\ \emph {et~al.}(2013)\citenamefont
  {Wojde{\l}}, \citenamefont {Hermet}, \citenamefont {Ljungberg}, \citenamefont
  {Ghosez},\ and\ \citenamefont {{\'I}{\~n}iguez}}]{WojdeJPCM13}%
  \BibitemOpen
\bibfield  {journal} {  }\bibfield  {author} {\bibinfo {author} {\bibfnamefont
  {J.~C.}\ \bibnamefont {Wojde{\l}}}, \bibinfo {author} {\bibfnamefont
  {P.}~\bibnamefont {Hermet}}, \bibinfo {author} {\bibfnamefont {M.~P.}\
  \bibnamefont {Ljungberg}}, \bibinfo {author} {\bibfnamefont {P.}~\bibnamefont
  {Ghosez}}, \ and\ \bibinfo {author} {\bibfnamefont {J.}~\bibnamefont
  {{\'I}{\~n}iguez}},\ }\href {\doibase 10.1088/0953-8984/25/30/305401}
  {\bibfield  {journal} {\bibinfo  {journal} {J. Phys. Condens. Matter}\
  }\textbf {\bibinfo {volume} {25}},\ \bibinfo {pages} {305401} (\bibinfo
  {year} {2013})}\BibitemShut {NoStop}%
\bibitem [{\citenamefont {Schelling}\ \emph {et~al.}(2002)\citenamefont
  {Schelling}, \citenamefont {Phillpot},\ and\ \citenamefont
  {Keblinski}}]{SchellingPRB02}%
  \BibitemOpen
  \bibfield  {author} {\bibinfo {author} {\bibfnamefont {P.~K.}\ \bibnamefont
  {Schelling}}, \bibinfo {author} {\bibfnamefont {S.~R.}\ \bibnamefont
  {Phillpot}}, \ and\ \bibinfo {author} {\bibfnamefont {P.}~\bibnamefont
  {Keblinski}},\ }\href@noop {} {\bibfield  {journal} {\bibinfo  {journal}
  {Phys. Rev. B}\ }\textbf {\bibinfo {volume} {65}},\ \bibinfo {pages} {144306}
  (\bibinfo {year} {2002})}\BibitemShut {NoStop}%
\bibitem [{\citenamefont {Escorihuela-Sayalero}\ \emph
  {et~al.}(2017)\citenamefont {Escorihuela-Sayalero}, \citenamefont
  {Wojde{\l}},\ and\ \citenamefont {{\'I}{\~n}iguez}}]{EscorihuelaPRB17}%
  \BibitemOpen
  \bibfield  {author} {\bibinfo {author} {\bibfnamefont {C.}~\bibnamefont
  {Escorihuela-Sayalero}}, \bibinfo {author} {\bibfnamefont {J.~C.}\
  \bibnamefont {Wojde{\l}}}, \ and\ \bibinfo {author} {\bibfnamefont
  {J.}~\bibnamefont {{\'I}{\~n}iguez}},\ }\href {\doibase
  10.1103/PhysRevB.95.094115} {\bibfield  {journal} {\bibinfo  {journal} {Phys.
  Rev. B}\ }\textbf {\bibinfo {volume} {95}},\ \bibinfo {pages} {094115}
  (\bibinfo {year} {2017})}\BibitemShut {NoStop}%
\bibitem [{\citenamefont {Wojde{\l}}\ and\ \citenamefont
  {{\'I}{\~n}iguez}(2014)}]{WojdePRL14}%
  \BibitemOpen
  \bibfield  {author} {\bibinfo {author} {\bibfnamefont {J.~C.}\ \bibnamefont
  {Wojde{\l}}}\ and\ \bibinfo {author} {\bibfnamefont {J.}~\bibnamefont
  {{\'I}{\~n}iguez}},\ }\href@noop {} {\bibfield  {journal} {\bibinfo
  {journal} {Phys. Rev. Lett.}\ }\textbf {\bibinfo {volume} {112}},\ \bibinfo
  {pages} {247603} (\bibinfo {year} {2014})}\BibitemShut {NoStop}%
\bibitem [{\citenamefont {Zubko}\ \emph {et~al.}(2016)\citenamefont {Zubko},
  \citenamefont {Wojde{\l}}, \citenamefont {Hadjimichael}, \citenamefont
  {Fernandez-Pena}, \citenamefont {Sen{\'e}}, \citenamefont {Luk'yanchuk},
  \citenamefont {Triscone},\ and\ \citenamefont {{\'I}{\~n}iguez}}]{Zubko2016}%
  \BibitemOpen
  \bibfield  {author} {\bibinfo {author} {\bibfnamefont {P.}~\bibnamefont
  {Zubko}}, \bibinfo {author} {\bibfnamefont {J.~C.}\ \bibnamefont
  {Wojde{\l}}}, \bibinfo {author} {\bibfnamefont {M.}~\bibnamefont
  {Hadjimichael}}, \bibinfo {author} {\bibfnamefont {S.}~\bibnamefont
  {Fernandez-Pena}}, \bibinfo {author} {\bibfnamefont {A.}~\bibnamefont
  {Sen{\'e}}}, \bibinfo {author} {\bibfnamefont {I.}~\bibnamefont
  {Luk'yanchuk}}, \bibinfo {author} {\bibfnamefont {J.-M.}\ \bibnamefont
  {Triscone}}, \ and\ \bibinfo {author} {\bibfnamefont {J.}~\bibnamefont
  {{\'I}{\~n}iguez}},\ }\href {\doibase 10.1038/nature17659} {\bibfield
  {journal} {\bibinfo  {journal} {Nature}\ }\textbf {\bibinfo {volume} {534}},\
  \bibinfo {pages} {524} (\bibinfo {year} {2016})}\BibitemShut {NoStop}%
\bibitem [{\citenamefont {Li}\ \emph {et~al.}(2014)\citenamefont {Li},
  \citenamefont {Carrete}, \citenamefont {Katcho},\ and\ \citenamefont
  {Mingo}}]{LiCPC14}%
  \BibitemOpen
  \bibfield  {author} {\bibinfo {author} {\bibfnamefont {W.}~\bibnamefont
  {Li}}, \bibinfo {author} {\bibfnamefont {J.}~\bibnamefont {Carrete}},
  \bibinfo {author} {\bibfnamefont {N.~A.}\ \bibnamefont {Katcho}}, \ and\
  \bibinfo {author} {\bibfnamefont {N.}~\bibnamefont {Mingo}},\ }\href@noop {}
  {\bibfield  {journal} {\bibinfo  {journal} {Comp. Phys. Commun.}\ }\textbf
  {\bibinfo {volume} {185}},\ \bibinfo {pages} {1747} (\bibinfo {year}
  {2014})}\BibitemShut {NoStop}%
\bibitem [{\citenamefont {Tachibana}\ \emph {et~al.}(2008)\citenamefont
  {Tachibana}, \citenamefont {Kolodiazhnyi},\ and\ \citenamefont
  {Takayama-Muromachi}}]{TachibanaAPL08}%
  \BibitemOpen
  \bibfield  {author} {\bibinfo {author} {\bibfnamefont {M.}~\bibnamefont
  {Tachibana}}, \bibinfo {author} {\bibfnamefont {T.}~\bibnamefont
  {Kolodiazhnyi}}, \ and\ \bibinfo {author} {\bibfnamefont {E.}~\bibnamefont
  {Takayama-Muromachi}},\ }\href {\doibase 10.1063/1.2978072} {\bibfield
  {journal} {\bibinfo  {journal} {Appl. Phys. Lett.}\ }\textbf {\bibinfo
  {volume} {93}},\ \bibinfo {pages} {092902} (\bibinfo {year}
  {2008})}\BibitemShut {NoStop}%
\bibitem [{\citenamefont {Sellan}\ \emph {et~al.}(2010)\citenamefont {Sellan},
  \citenamefont {Landry}, \citenamefont {Turney}, \citenamefont {McGaughey},\
  and\ \citenamefont {Amon}}]{SellanPRB10}%
  \BibitemOpen
  \bibfield  {author} {\bibinfo {author} {\bibfnamefont {D.~P.}\ \bibnamefont
  {Sellan}}, \bibinfo {author} {\bibfnamefont {E.~S.}\ \bibnamefont {Landry}},
  \bibinfo {author} {\bibfnamefont {J.~E.}\ \bibnamefont {Turney}}, \bibinfo
  {author} {\bibfnamefont {A.~J.~H.}\ \bibnamefont {McGaughey}}, \ and\
  \bibinfo {author} {\bibfnamefont {C.~H.}\ \bibnamefont {Amon}},\ }\href@noop
  {} {\bibfield  {journal} {\bibinfo  {journal} {Phys. Rev. B}\ }\textbf
  {\bibinfo {volume} {81}},\ \bibinfo {pages} {214305} (\bibinfo {year}
  {2010})}\BibitemShut {NoStop}%
\bibitem [{\citenamefont {Rurali}\ \emph {et~al.}(2016)\citenamefont {Rurali},
  \citenamefont {Colombo}, \citenamefont {Cartoix\`a}, \citenamefont
  {Wilhelmsen}, \citenamefont {Trinh}, \citenamefont {Bedeaux},\ and\
  \citenamefont {Kjelstrup}}]{RuraliPCCP16}%
  \BibitemOpen
  \bibfield  {author} {\bibinfo {author} {\bibfnamefont {R.}~\bibnamefont
  {Rurali}}, \bibinfo {author} {\bibfnamefont {L.}~\bibnamefont {Colombo}},
  \bibinfo {author} {\bibfnamefont {X.}~\bibnamefont {Cartoix\`a}}, \bibinfo
  {author} {\bibfnamefont {O.}~\bibnamefont {Wilhelmsen}}, \bibinfo {author}
  {\bibfnamefont {T.~T.}\ \bibnamefont {Trinh}}, \bibinfo {author}
  {\bibfnamefont {D.}~\bibnamefont {Bedeaux}}, \ and\ \bibinfo {author}
  {\bibfnamefont {S.}~\bibnamefont {Kjelstrup}},\ }\href@noop {} {\bibfield
  {journal} {\bibinfo  {journal} {Phys. Chem. Chem. Phys.}\ }\textbf {\bibinfo
  {volume} {18}},\ \bibinfo {pages} {13741} (\bibinfo {year}
  {2016})}\BibitemShut {NoStop}%
\bibitem [{px()}]{px}%
  \BibitemOpen
  \href@noop {} {}\bibinfo {howpublished} {The spontaneous polarization $P_x$,
  that defines the ferroelectric ground state, might appear to be a sounder
  choice. However, it varies more abruptly than $P_y$, thus we believe the
  latter better capture the non-bulk nature of a layer of material. Taking the
  variation $P_x$ to quantify the intreface thickness we obtain
  12~\AA.}\BibitemShut {Stop}%
\bibitem [{\citenamefont {Kjelstrup}\ and\ \citenamefont
  {Bedeaux}(2008)}]{SigneDick}%
  \BibitemOpen
  \bibfield  {author} {\bibinfo {author} {\bibfnamefont {S.}~\bibnamefont
  {Kjelstrup}}\ and\ \bibinfo {author} {\bibfnamefont {D.}~\bibnamefont
  {Bedeaux}},\ }in\ \href@noop {} {\emph {\bibinfo {booktitle}
  {{Non-Equilibrium Thermodynamics of Heterogeneous Systems}}}},\ Vol.~\bibinfo
  {volume} {16}\ (\bibinfo  {publisher} {World Scientific},\ \bibinfo {address}
  {Singapore},\ \bibinfo {year} {2008})\BibitemShut {NoStop}%
\bibitem [{\citenamefont {Dettori}\ \emph {et~al.}(2016)\citenamefont
  {Dettori}, \citenamefont {Melis}, \citenamefont {Cartoix\`a}, \citenamefont
  {Rurali},\ and\ \citenamefont {Colombo}}]{DettoriAiPX16}%
  \BibitemOpen
  \bibfield  {author} {\bibinfo {author} {\bibfnamefont {R.}~\bibnamefont
  {Dettori}}, \bibinfo {author} {\bibfnamefont {C.}~\bibnamefont {Melis}},
  \bibinfo {author} {\bibfnamefont {X.}~\bibnamefont {Cartoix\`a}}, \bibinfo
  {author} {\bibfnamefont {R.}~\bibnamefont {Rurali}}, \ and\ \bibinfo {author}
  {\bibfnamefont {L.}~\bibnamefont {Colombo}},\ }\href@noop {} {\bibfield
  {journal} {\bibinfo  {journal} {Adv. Phys. X}\ }\textbf {\bibinfo {volume}
  {1}},\ \bibinfo {pages} {246} (\bibinfo {year} {2016})}\BibitemShut {NoStop}%
\bibitem [{\citenamefont {Pourfath}(2014)}]{PourfathBOOK14}%
  \BibitemOpen
  \bibfield  {author} {\bibinfo {author} {\bibfnamefont {M.}~\bibnamefont
  {Pourfath}},\ }\href@noop {} {\emph {\bibinfo {title} {{The Non-Equilibrium
  Green's Function Method for Nanoscale Device Simulation}}}}\ (\bibinfo
  {publisher} {Springer},\ \bibinfo {year} {2014})\BibitemShut {NoStop}%
\bibitem [{\citenamefont {Sadasivam}\ \emph {et~al.}(2014)\citenamefont
  {Sadasivam}, \citenamefont {Che}, \citenamefont {Huang}, \citenamefont
  {Chen}, \citenamefont {Kumar},\ and\ \citenamefont
  {Fisher}}]{SadasivamARHT14}%
  \BibitemOpen
  \bibfield  {author} {\bibinfo {author} {\bibfnamefont {S.}~\bibnamefont
  {Sadasivam}}, \bibinfo {author} {\bibfnamefont {Y.}~\bibnamefont {Che}},
  \bibinfo {author} {\bibfnamefont {Z.}~\bibnamefont {Huang}}, \bibinfo
  {author} {\bibfnamefont {L.}~\bibnamefont {Chen}}, \bibinfo {author}
  {\bibfnamefont {S.}~\bibnamefont {Kumar}}, \ and\ \bibinfo {author}
  {\bibfnamefont {T.~S.}\ \bibnamefont {Fisher}},\ }\href@noop {} {\bibfield
  {journal} {\bibinfo  {journal} {Ann. Rev. Heat Transfer}\ }\textbf {\bibinfo
  {volume} {17}},\ \bibinfo {pages} {89} (\bibinfo {year} {2014})}\BibitemShut
  {NoStop}%
\bibitem [{\citenamefont {Sancho}\ \emph {et~al.}(1984)\citenamefont {Sancho},
  \citenamefont {Sancho},\ and\ \citenamefont {Rubio}}]{SanchoJPF84}%
  \BibitemOpen
  \bibfield  {author} {\bibinfo {author} {\bibfnamefont {M.~L.}\ \bibnamefont
  {Sancho}}, \bibinfo {author} {\bibfnamefont {J.~L.}\ \bibnamefont {Sancho}},
  \ and\ \bibinfo {author} {\bibfnamefont {J.}~\bibnamefont {Rubio}},\
  }\href@noop {} {\bibfield  {journal} {\bibinfo  {journal} {J. Phys. F: Met.
  Phys.}\ }\textbf {\bibinfo {volume} {14}},\ \bibinfo {pages} {1205} (\bibinfo
  {year} {1984})}\BibitemShut {NoStop}%
\bibitem [{\citenamefont {Lu}\ and\ \citenamefont {McGaughey}(2015)}]{LuAIP15}%
  \BibitemOpen
  \bibfield  {author} {\bibinfo {author} {\bibfnamefont {S.}~\bibnamefont
  {Lu}}\ and\ \bibinfo {author} {\bibfnamefont {A.~J.~H.}\ \bibnamefont
  {McGaughey}},\ }\href {\doibase 10.1063/1.4918591} {\bibfield  {journal}
  {\bibinfo  {journal} {AIP Advances}\ }\textbf {\bibinfo {volume} {5}},\
  \bibinfo {pages} {053205} (\bibinfo {year} {2015})}\BibitemShut {NoStop}%
\end{thebibliography}%

\newpage

\newpage
\clearpage

\newpage
\clearpage

\newpage
\clearpage

\newpage
\clearpage

\end{document}